\documentclass[runningheads]{llncs}
\usepackage[T1]{fontenc}
\usepackage{CJKutf8}
\usepackage{graphicx}
\usepackage{amsmath}
\usepackage{amssymb}
\usepackage{hyperref}
\usepackage{booktabs}
\usepackage{listings}
\usepackage{xcolor}
\usepackage[section]{placeins}
\usepackage{float}

\begin{document}

\title{NormCode Canvas: Making LLM Agentic Workflows Development Sustainable via Case-Based Reasoning}

\titlerunning{NormCode Canvas: Making LLM Agentic Workflows Sustainable via CBR}

\author{Xin Guan\inst{1}\orcidID{0000-0003-1670-4937} \and
Yunshan Li\inst{2,1} \and
Ze Wang\inst{3,1}}
\authorrunning{Guan et al.}

\institute{Psylens AI, China \and
Shenzhen University, China \and
University College London, United Kingdom\\
\email{garguan2001@outlook.com}}

\maketitle

\begin{abstract}
We present \textbf{NormCode Canvas} (v1.1.3), a deployed system realizing
Case-Based Reasoning at two levels for multi-step LLM workflows. The foundation
is NormCode \cite{guan2026normcodesemiformallanguageauditable}, a semi-formal planning language whose compiler-verified
scope rule ensures every execution checkpoint is a genuinely self-contained case ---
eliminating the implicit shared state that makes retrieval unreliable and failure
non-localizable in standard orchestration frameworks.
\textbf{Level~1} treats each checkpoint as a concrete case (suspended runtime);
Fork implements retrieve-and-reuse, Value Override implements revision with automatic
stale-boundary propagation.
\textbf{Level~2} treats each compiled plan as an abstract case; the compilation
pipeline is itself a NormCode plan, enabling recursive case learning.
Three structural properties follow: \textbf{(C1)} direct checkpoint
inspection; \textbf{(C2)} pre-execution review via compiler-generated narrative;
\textbf{(C3)} scope-bounded selective re-execution.
Four deployed plans serve as structured evidence: PPT Generation produces
presentation decks at ${\sim}40$\,s per slide on commercial APIs; Code Assistant
carries out multi-step software-engineering tasks spanning up to ten reasoning
cycles; NC~Compilations converts natural-language specifications into executable
NormCode plans; and Canvas Assistant, when connected to an external AI code editor,
automates plan debugging. Together these plans form a self-sustaining ecosystem in
which plans produce, debug, and refine one another --- realizing cumulative
case-based learning at system scale.

\end{abstract}

\section{Introduction}
\label{sec:intro}

Multi-step LLM workflows---where each inference step feeds the next---have emerged as a
standard architecture for deploying language models in real applications. Frameworks such
as LangChain \cite{chase2022}, LangGraph \cite{langgraph2024}, AutoGen \cite{wu2023},
and PromptFlow \cite{microsoft2023} have made such workflows easy to build. Yet they
remain difficult to debug, resume, and improve systematically: state is carried forward
through conversation histories, mutable objects, and framework-specific control logic,
so an intermediate checkpoint at step $k$ may silently depend on earlier context not
captured in the checkpoint itself. Failure recovery becomes brittle, replay hard to
trust, and debugging requires reconstructing large portions of prior execution.

Tooling can add persistence and tracing, but the underlying limitation remains: the
plan, its intermediate semantics, and the conditions for valid continuation are scattered
across code and prompts rather than expressed in a single interpretable form.

\textbf{What is missing is a workflow language} in which multi-step reasoning can be
explicitly represented, inspected, verified, executed, and revised --- simultaneously
human-readable, machine-executable, and structurally enforceable.

\textbf{NormCode} \cite{guan2026normcodesemiformallanguageauditable} is a semi-formal planning language designed for that
role. Its scope rule guarantees that every execution checkpoint is a
\textbf{self-contained case} with no implicit context dependency
(Sect.~\ref{sec:normcode}). Three practical capabilities follow from this guarantee:
direct inspection of the exact scoped inputs at any flow index~(C1), pre-execution
structural review of the compiled plan~(C2), and scope-bounded selective
re-execution~(C3). \textbf{NormCode Canvas~(v1.1.3)} is the deployed system that
realizes C1--C3 and exposes a two-level CBR architecture: Level~1 manages concrete
cases (suspended runtimes at each checkpoint); Level~2 manages abstract cases
(executable plans that encode reusable workflow patterns). We do \textit{not} claim
universal superiority in end-task quality; our claims concern inspectability,
revisability, and bounded re-execution of multi-step LLM workflows.

\textbf{Relation to prior work.} The prior work~\cite{guan2026normcodesemiformallanguageauditable} is a
language specification: it defines NormCode's grammar, type system, and scope rule.
That prior work answers \textit{what NormCode is}. This paper answers \textit{what NormCode
makes possible in practice}, grounded in four deployed workflows ranging from PPT
generation and code assistance to self-hosted compilation and interactive canvas
control.

\textbf{Contributions.}
(1)~A two-level CBR account of LLM workflow development
(Sect.~\ref{sec:level1}--\ref{sec:level2}): Level~1 cases as suspended runtimes,
Level~2 cases as executable plans, and a compilation pipeline that is itself a
Level~2 case.
(2)~A scoped workflow language satisfying six properties
(Sect.~\ref{sec:normcode}): Interpretable, Enforceable, Composable, Locally
Addressable, Generalizable, and Portable.
(3)~A deployed system realizing three structural capabilities
(Sect.~\ref{sec:canvas}--\ref{sec:casestudies}): direct checkpoint inspection~(C1),
pre-execution review~(C2), and selective re-execution~(C3).
(4)~Structured evidence from four production workflows
(Sect.~\ref{sec:casestudies}), including an ecosystem where plans produce, debug,
and refine one another.

\section{Background}
\label{sec:background}

\subsection{CBR and Process-Oriented CBR}

Case-Based Reasoning \cite{aamodt1994,kolodner1993} solves new problems by retrieving
and adapting past cases. The four-stage cycle --- \textbf{Retrieve}, \textbf{Reuse},
\textbf{Revise}, \textbf{Retain} --- depends critically on case integrity: each case
must be a self-contained record of a past solution state, free of implicit context.

Process-Oriented CBR (POCBR) \cite{bergmann2002,minor2014} extends this to
\textbf{processes as cases}: the case captures a workflow trace including intermediate
states and data flow. Multi-step LLM workflows are a natural POCBR domain; the
challenge is ensuring their intermediate states are suitable, self-contained cases.

\subsection{Case Integrity and Structural Isolation}

Most LLM orchestration frameworks do not satisfy case integrity by default: a
checkpoint at step $k$ may embed accumulated conversation history or prompt context
from steps $1, \ldots, k-1$, making it a snapshot of shared state rather than a
self-contained case. Three consequences follow: \textbf{unreliable retrieval} ---
continuing from such a checkpoint in a new context produces incorrect behavior;
\textbf{non-localizable failures} --- the cause of a failure at step $k$ may be
anywhere in accumulated implicit state; \textbf{irreproducible cases} --- two
executions loading the same checkpoint may produce different outputs.

These failures are eliminated by \textbf{structural isolation}: a workflow language
provides structural isolation if each step can access only data explicitly declared in
its own scope block, the compiler determines each step's scope at formalization, and the
orchestrator constructs each step's input from only its declared references. Under this
guarantee, the stored output at step $k$ contains exactly what downstream steps need;
failures are localizable to the declared inputs of the failing step; and checkpoint
reuse reproduces the original data-dependency behavior. NormCode \cite{guan2026normcodesemiformallanguageauditable} is
designed around this enabling condition.

\section{NormCode as a Workflow Language}
\label{sec:normcode}

NormCode is a representational language in which reasoning plans are formulated, not
merely described. This section argues that NormCode satisfies six properties needed of a
practical workflow language for AI orchestration. The formal language specification is
in \cite{guan2026normcodesemiformallanguageauditable}. Prior to activation, a NormCode plan exists in three forms: the human-authored
semi-formal source (\texttt{.ncds}), the rigorously formalized compiled form
(\texttt{.ncd}), and the plain-prose narrative (\texttt{.ncn}) generated for
pre-execution review.

\subsection{Six Properties of a Practical Workflow Language}

(P1)~\textbf{Interpretable}: the \texttt{.ncds} format reads like structured natural
language; Formalization produces \texttt{.ncn}, a plain-prose narrative for
pre-execution review (C2); concept names are natural-language identifiers, so plans can
be authored in any language. (P2)~\textbf{Enforceable}: the compiler determines each
step's scope before the first run; the \texttt{.ncn} is a deterministic rendering of
the actual formal plan, not a paraphrase. (P3)~\textbf{Composable}: three symbols and
one scope rule express nested loops, conditionals, parallel derivation, and multi-agent
coordination at any plan scale. (P4)~\textbf{Locally Addressable}: every step has a
dot-separated flow index; each step is self-contained, so any checkpoint can be
examined, revised, or reused without reconstructing preceding steps.
(P5)~\textbf{Generalizable}: the same three primitives support multiple workflow classes
in our deployed examples, including PPT generation, code assistance, self-hosted
compilation, and interactive canvas control, without domain-specific syntax.
(P6)~\textbf{Portable}: a NormCode plan is a plain-text
\texttt{.ncds} with \texttt{concept\_repo.json} and \texttt{inference\_repo.json} ---
version-controlled, shareable, and executable by any compliant orchestrator.

\subsection{Syntax and the Scope Rule}

\begin{figure}[!htbp]
  \centering
  \includegraphics[width=0.88\textwidth]{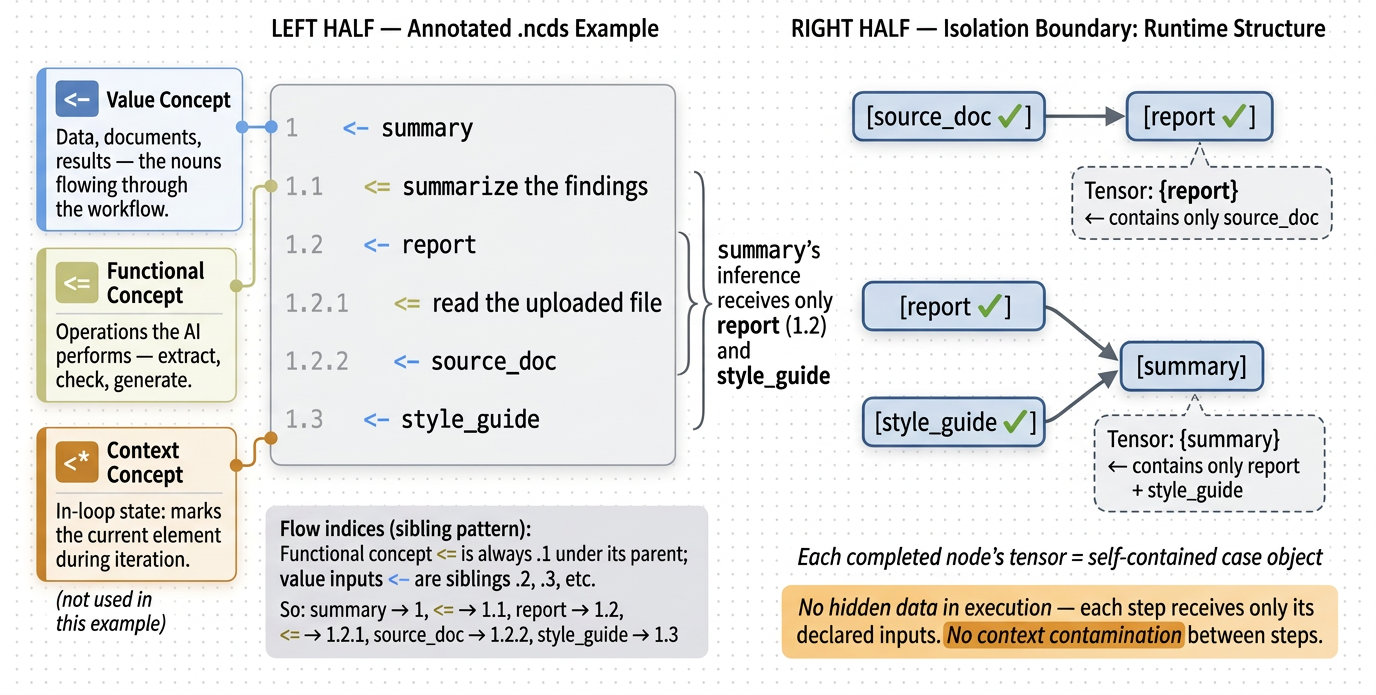}
  \caption{NormCode syntax and scope rule. Left: annotated \texttt{.ncds}; right:
    isolation boundary at runtime. Each node's tensor is self-contained (C1).}
  \label{fig:scope}
\end{figure}

Three markers constitute the full syntax (Fig.~\ref{fig:scope}; formal specification
in \cite{guan2026normcodesemiformallanguageauditable}):
\texttt{<-}~Value, \texttt{<=}~Functional, \texttt{<*}~Context (loop-current element).
The language is \textbf{semi-formal}: concept names and descriptions are free-form
natural-language strings (e.g.\ \texttt{\{document summary\}} or
\texttt{\{}\begin{CJK}{UTF8}{gbsn}分析\end{CJK}\texttt{\}}) while the three
markers and indentation are rigidly enforced, making plans simultaneously readable
by domain experts, compilable by the orchestrator, and
\textbf{fully language-agnostic} (Sect.~\ref{sec:casestudies}). Each inference's inputs are exactly the concepts
declared in its immediately enclosing block --- no implicit sharing. The compiler
determines scope at formalization; the orchestrator enforces it at runtime.
Plans are read top-to-bottom (root concept first) and execute inside-out (deepest
dependencies resolved first). Flow indices (\texttt{run\_id}, \texttt{flow\_index})
give every checkpoint a stable address. \textbf{Semantic inferences} invoke LLMs
(probabilistic, token-consuming); \textbf{syntactic inferences} perform deterministic
data manipulation (routing, grouping, timing, looping) at zero LLM cost --- 60--70\%
of nodes in production plans. Syntactic checkpoints are perfectly reproducible;
semantic checkpoints are bounded-stochastic. Concept data are stored as
\textbf{References} (named-axis tensors); large inputs pass as \textit{perceptual
signs} --- compact pointers transmuted to content only at the moment of an LLM call,
preventing context accumulation. The Detail Panel exposes every completed step's exact
bounded inputs with constant-effort direct inspection (failure localization, C1).

\section{Two-Level CBR: Level 1 --- Concrete Case Management}
\label{sec:level1}

\begin{figure}[!htbp]
  \centering
  \includegraphics[width=0.92\textwidth]{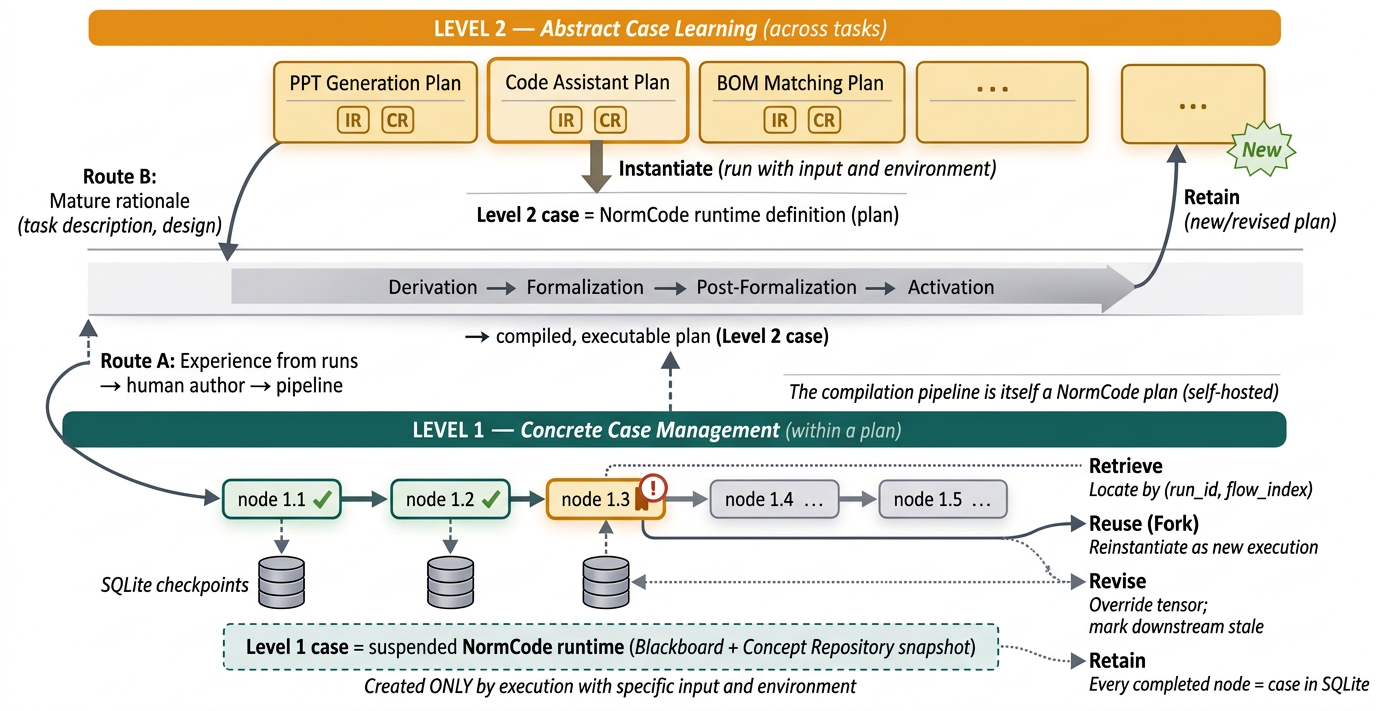}
  \caption{Two-level CBR architecture. Level~2 (top): abstract cases (plans);
    Level~1 (bottom): concrete cases (suspended runtimes). Route~A distills
    experience into plans; Route~B instantiates plans to generate Level~1 cases.}
  \label{fig:architecture}
\end{figure}

\subsection{Level 1 Case Definition and Case Base}

Fig.~\ref{fig:architecture} illustrates the two-level architecture.
A Level~1 case is a \textbf{suspended NormCode runtime} --- the full execution
environment at a specific flow index, captured and persisted. Formally, a Level~1 case
$c = (r,\; f,\; \mathcal{B}_f,\; \mathcal{C}_f,\; \mathcal{E}_f^{\,?})$, where $r$ is
the run identifier; $f$ is the flow index of the last completed node; $\mathcal{B}_f$
is the \textbf{Blackboard snapshot}: the full node status map at the moment $f$
completed; $\mathcal{C}_f$ is the \textbf{Concept Repository snapshot}: all tensors for
all nodes completed through $f$; $\mathcal{E}_f^{\,?}$ is an optional environment
snapshot, present when $\mathcal{C}_f$ contains perceptual signs. Because the scope
rule prevents any node from accessing data outside its declared block, $\mathcal{C}_f$
is self-contained --- there is no hidden context dependency. The \textit{case base}
$\mathcal{CB}$ is the set of all Level~1 cases across all runs:
\begin{equation}
  \mathcal{CB} = \{(r, f, \mathcal{B}_f, \mathcal{C}_f, \mathcal{E}_f^{\,?}) \mid
  r \in \text{Runs},\; f \in \text{FlowIndices}(r)\}
\end{equation}
$\mathcal{CB}$ is persisted in a per-project SQLite database; the Checkpoint Panel is
its visual interface.

\textbf{Retrieve.} Retrieval selects a past case from $\mathcal{CB}$ as the starting
point for a new execution. Two triggers: (1)~a run fails at flow index $f^*$;
(2)~a new task is similar to a past execution. Case inspection via the Tensor Inspector
opens $\mathcal{C}_f$ at any flow index in table, list, or JSON view. Because
$\mathcal{C}_f$ contains exactly the bounded inputs of each completed step, this
inspection is direct and constant-effort --- failure localization (C1).

\textbf{Reuse --- Fork.} The Fork operation implements reuse: (1)~new run identifier
$r'$ is created; (2)~orchestrator loads $(\mathcal{B}_f, \mathcal{C}_f)$ from retrieved
case; (3)~Blackboard marks all nodes at flow indices $\leq f$ as completed;
(4)~execution proceeds from the first pending node. No upstream re-execution.

\textbf{Revise --- Value Override + Selective Re-run.} (1)~Operator retrieves a case at
flow index $f$; (2)~overrides a tensor in $\mathcal{C}_f$ via the Tensor Inspector edit
mode; (3)~orchestrator marks all downstream nodes as stale; (4)~on resume, only stale
nodes re-execute; (5)~revised run produces new cases at every completed node. Because
the scope rule guarantees that only nodes downstream of $f$ can depend on the tensor at
$f$, the stale set is exactly determined by the data dependency graph --- selective
re-execution (C3). No manual specification required.

\textbf{Retain.} Every node completed during a run writes a new case to $\mathcal{CB}$
automatically. The case base accumulates all runs' Blackboard and Concept Repository
snapshots continuously.

\section{Two-Level CBR: Level 2 --- Abstract Case Learning}
\label{sec:level2}

Level~1 CBR (Sect.~\ref{sec:level1}) operates \textit{within} a plan. Level~2 CBR
operates \textit{across} plans: it treats the plans themselves as cases --- abstract
workflow patterns that encode a reusable solution structure for a class of problems.

\subsection{What a Level 2 Case Is}

A Level~2 case is a \textbf{NormCode runtime definition}: an executable plan that
specifies the reasoning structure for a task type. Plans may be authored manually,
produced through the compilation pipeline (Sect.~\ref{sec:level2:distillation}), or
developed through iterative human-AI collaboration --- the defining property is that
the result is a fully executable \texttt{.ncds} + repositories artifact, not the path
by which it was produced.

\textbf{Definition.} A \textit{Level~2 case} $P$ is a triple:
\begin{equation}
  P = (\mathcal{IR},\; \mathcal{CR},\; \mathcal{V})
\end{equation}
where $\mathcal{IR}$ is the \textbf{Inference Repository}: the complete inference graph
with execution configurations (paradigm assignments, body faculties, perception norms);
$\mathcal{CR}$ is the \textbf{Concept Repository}: concept definitions with tensor axes,
shapes, and ground references (perceptual signs pointing to input resources);
$\mathcal{V}$ is the \textbf{Provision set}: the invariant vertical resources --- prompt
templates, scripts, paradigm files, and configuration parameters --- bound at
compilation time and referenced as ground inputs across all runs of this plan. $\mathcal{V}$
is what distinguishes an abstract case from a bare inference graph: it encodes
\textit{how} operations execute, not merely their structure.

\subsection{The Level 2 CBR Cycle}

\textbf{Retrieve}: a practitioner reads the \texttt{.ncn} narrative and determines
whether an existing plan (or part of it) matches the new task class.
\textbf{Reuse}: the retrieved plan is instantiated with new input values and run; all
four production-ready plans are reused in this sense on every run, each generating a
fresh set of Level~1 cases. \textbf{Revise}: when the plan does not suit the new task,
the designer modifies the \texttt{.ncds} source and recompiles; the \texttt{.ncn}
re-review verifies the revised abstract case before it generates any concrete cases.
\textbf{Retain}: the revised or newly authored plan joins the case library, which grows
through use.

\subsection{The Compilation Pipeline for NormCode Production}
\label{sec:level2:distillation}

The compilation pipeline (Derivation $\rightarrow$ Formalization $\rightarrow$
Post-Formalization $\rightarrow$ Activation) is one standardized distillation path that
produces Level~2 cases from experience and intent: \textit{Derivation} extracts
conceptual structure from natural language; \textit{Formalization} assigns flow indices
and determines execution types; \textit{Post-Formalization} assigns execution resources;
\textit{Activation} serializes the abstract case into orchestrator-ready JSON.

A notable consequence of NormCode as a workflow language is that \textbf{the
compilation pipeline itself is a NormCode plan.} This pipeline is a Level~2 case
(the NC~Compilations plan), but it is not a universal compiler. What it produces
depends on the task's structure and on the quality of the input specification. For
plans with clear sequential structure, the pipeline can reliably produce executable
plans. For plans with richer domain logic, heavier tool interaction, or more open-ended
control flow, manual authoring or human-assisted iteration remains the practical path.
The NC~Compilations \texttt{.ncds} is shown alongside the other production plans in
Fig.~\ref{fig:plans}.

The recursive structure has three consequences: (1)~\textbf{Validation} --- the pipeline
is self-consistent and can compile itself; (2)~\textbf{Improvement through revision} ---
improving the compilation pipeline is abstract case revision at the meta-level;
(3)~\textbf{Recursive retention} --- each improvement enriches the case library. Scope
and limits of automated compilation are discussed in Sect.~\ref{sec:discussion}.

\section{NormCode Canvas: The Deployed System}
\label{sec:canvas}

\begin{figure}[!htbp]
  \centering
  \includegraphics[width=\textwidth]{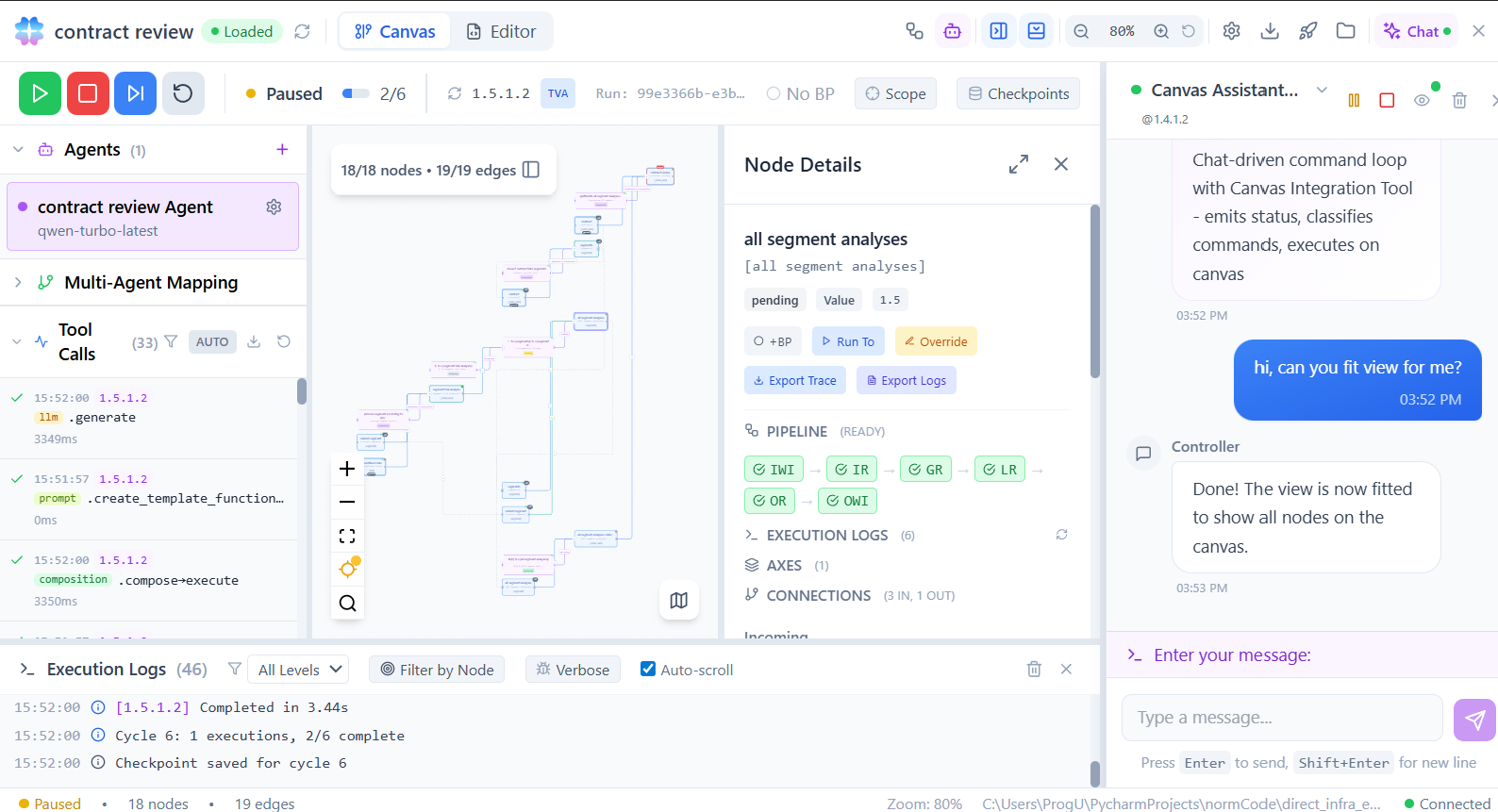}
  \caption{NormCode Canvas (v1.1.3). The Graph Canvas (center) renders the full
    inference graph; the Detail Panel (right) inspects any node's bounded inputs ---
    failure localization (C1).}
  \label{fig:canvas}
\end{figure}

NormCode Canvas (v1.1.3) is the deployed system that makes both levels of CBR
accessible in one environment (Fig.~\ref{fig:canvas}). Three practical properties
follow structurally from the underlying workflow language and are realized by Canvas:
(1)~\textbf{(C1) Direct checkpoint inspection.} At any breakpoint, the exact bounded
inputs to step $k$ are directly inspectable --- not reconstructed from logs --- because
the scope rule enforced them at runtime. (2)~\textbf{(C2) Pre-execution
structural review.} The \texttt{.ncn} narrative is available before any LLM call; a
domain expert can verify and reject the plan before computation begins. Unlike RLHF
\cite{ouyang2022} or Constitutional AI \cite{bai2022}, this intervention operates at
the plan structure level. (3)~\textbf{(C3) Scope-bounded selective re-execution.}
Revising a case at flow index $f$ triggers re-execution of only transitively downstream
nodes; all upstream results are reused from cache.

\subsection{Level 2 Interface: Abstract Case Management}

A Level~2 case in Canvas is a \textbf{NormCode project}: a directory containing the
\texttt{.ncds} source, compiled \texttt{concept\_repo.json} and
\texttt{inference\_repo.json}, agent configuration, and execution settings. The Project
Panel is the entry point for opening, creating, and switching between projects. Before
the first run, the \texttt{.ncn} narrative is available for pre-execution review (C2):
a domain expert reads the plain-prose rendering, verifies that the reasoning structure
matches intent, and approves or rejects before any LLM call. Canvas additionally ships
a built-in \textbf{Canvas Assistant} NormCode plan that runs inside the app, dispatching
natural-language messages to Canvas Integration facilities (\texttt{me.hands},
\texttt{me.mind}, \texttt{me.vision}); its per-iteration checkpoints are themselves
Level~1 cases. When the NC~Compilations plan runs, each compilation phase is a Level~1
node --- making the compilation pipeline debuggable via Level~1 CBR.

\subsection{Level 1 Interface: Concrete Case Management}

Level~1 cases are stored in a \textbf{per-project SQLite database}. Every completed
node writes a checkpoint automatically (\texttt{run\_id}, \texttt{flow\_index},
Blackboard snapshot, Concept Repository snapshot) --- Retain is fully automatic. The
Graph Canvas renders the full inference graph before execution; every node is a
potential checkpoint identified by its flow index. \textbf{Retrieve}: any node can be
designated as a breakpoint; execution pauses and the Detail Panel displays the node's
full $N$-dimensional Reference directly --- failure localization (C1). \textbf{Revise}:
the operator clicks Override at any breakpoint or completed case, edits a tensor, and
resumes; only downstream stale nodes re-execute --- selective re-execution (C3).
\textbf{Reuse}: the Checkpoint Panel browses all runs within the active project; Fork
loads $(\mathcal{B}_f, \mathcal{C}_f)$ into a new run $r'$, marks upstream nodes
completed, and executes downstream fresh --- supporting failure recovery, A/B testing,
and iterative refinement.

\subsection{Trace and Debugging Interface}

\begin{figure}[!htbp]
  \centering
  \includegraphics[width=\textwidth]{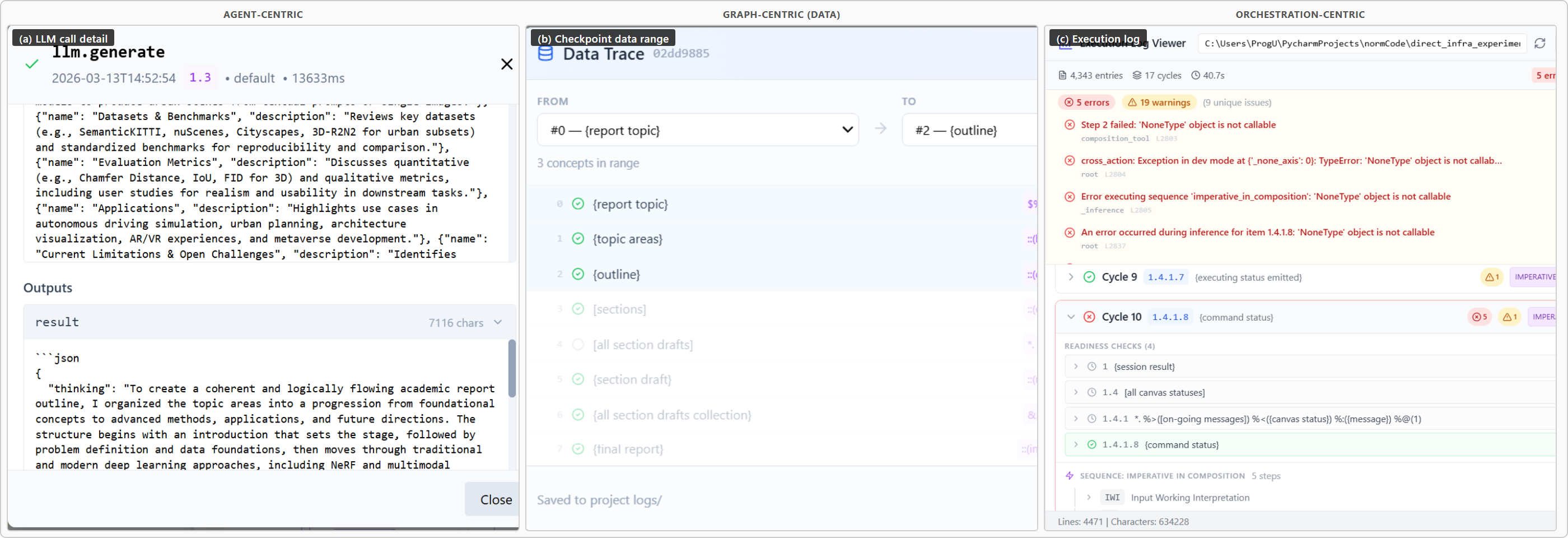}
  \caption{Three trace views in NormCode Canvas. (a)~Agent-centric: LLM call detail
    with prompt input and structured output. (b)~Graph-centric: scoped reference data
    between selected node indices~(C1). (c)~Orchestration-centric: execution log with
    error chain and cycle-level breakdown.}
  \label{fig:traces}
\end{figure}

Canvas exposes three complementary trace views (Fig.~\ref{fig:traces}).
The \textbf{agent-centric trace} records every tool call and its outcome ---
perception, prompt assembly, composition, and the LLM call --- with timestamps,
flow indices, and durations; expanding an entry reveals the full prompt input and
the model's structured output (Fig.~\ref{fig:traces}\,a).
The \textbf{graph-centric data trace} selects a range of nodes by flow index and
displays the scoped reference data that flowed between them
(Fig.~\ref{fig:traces}\,b). Because the scope rule guarantees self-contained
inputs, this view is a direct realisation of checkpoint inspection~(C1): the
operator sees exact tensor content without reconstructing it from logs.
The \textbf{orchestration-centric trace} logs every scheduling decision, readiness
check, error chain, and cycle-level breakdown (Fig.~\ref{fig:traces}\,c) ---
the view operators turn to for syntactical or structural failures that originate
inside the orchestrator rather than in the model's output.
In practice, the agent and data traces are checked first as they resolve the most
common failures; the orchestration trace is consulted when finer-grained evidence
is needed.

\subsection{Architecture}

\textbf{Frontend:} React~18 with TypeScript; React Flow for graph visualization;
TailwindCSS. \textbf{Backend:} FastAPI with Python~3.11; per-project SQLite for
Level~1 case persistence; Python WebSockets for real-time execution events.
\textbf{Orchestrator:} Dependency-driven scheduling via Waitlist and Blackboard; the
scope rule is enforced by construction --- only declared tensor references are passed to
each inference, never the full Concept Repository. Multiple agents with different LLM
models (\texttt{qwen-plus}, \texttt{gpt-4o}, \texttt{claude-3-opus}) can be
registered; inferences are mapped to agents by pattern rules on flow indices.

\section{Case Studies}
\label{sec:casestudies}

We use four deployed NormCode plans as \textbf{structured evidence} for the paper's
claims, not as a benchmark of universal output quality. Together they cover distinct
workflow classes and show that C1--C3 hold across different control-flow patterns,
human review points, and revision boundaries. Syntactic nodes constitute 60--70\% of
total inferences across these plans. NC~Compilations has been executed end-to-end with
\texttt{qwen-plus}, producing correct repositories for plans such as report writing and
contract review. Two of the four plans are then traced in more detail below.

\begin{figure}[!htbp]
  \centering
  \includegraphics[width=\textwidth]{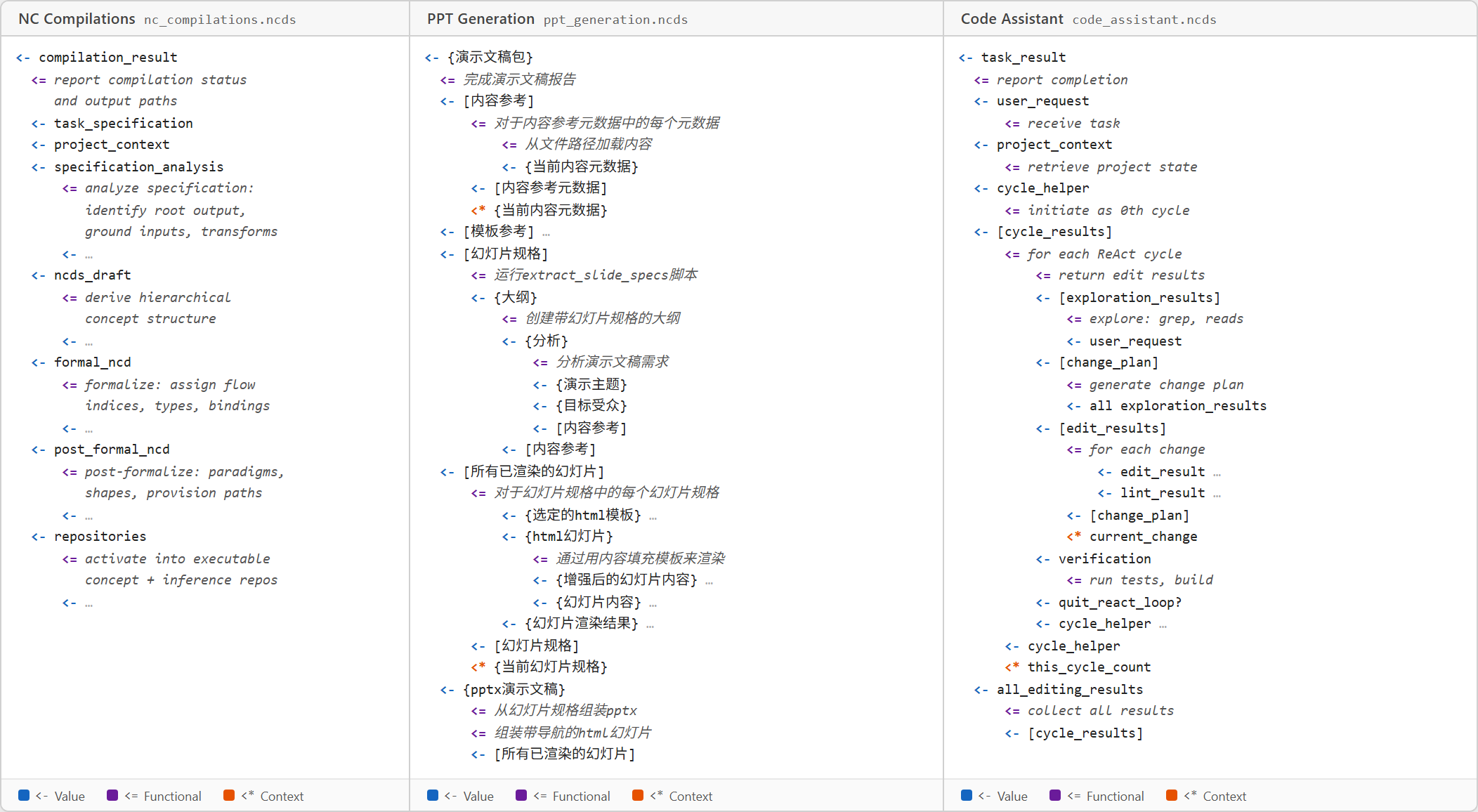}
  \caption{Core excerpts from three production NormCode plans as Level~2 cases
    (\texttt{.ncds}); actual production plans are larger and more complex.
    Left: NC~Compilations --- sequential compilation pipeline
    (Sect.~\ref{sec:level2:distillation}). Center: PPT Generation --- nested
    dual-output loop (HTML + PPTX per slide). Right: Code Assistant --- ReAct
    outer loop \cite{yao2023} with inner explore, plan, edit, and verify stages.
    \textcolor{blue}{\texttt{<-}} Value; \textcolor{violet}{\texttt{<=}} Functional;
    \textcolor{orange}{\texttt{<*}} Context. All three plans are reviewable as
    explicit plans before any run begins (C2).}
  \label{fig:plans}
\end{figure}

\subsection{Case Study 1: PPT Agent --- Level 2 Reuse and Level 1 Accumulation}

\texttt{ppt\_generation.ncds} (Fig.~\ref{fig:plans}, center) is a Level~2 case encoding
a five-phase pipeline: (0)~load reference files, (1)~analyse requirements and generate
a slide outline, (2)~extract per-slide specifications via script, (3)~a per-slide loop
producing both an HTML slide and a PPTX specification from the same componentized
content JSON, and (4)~assemble the final presentation. The plan employs a
\textbf{component-based design} with 6~layouts and 14~component types combined per
slide, so a single structured content JSON drives both renderers.

\begin{figure}[H]
  \centering
  \includegraphics[width=\textwidth]{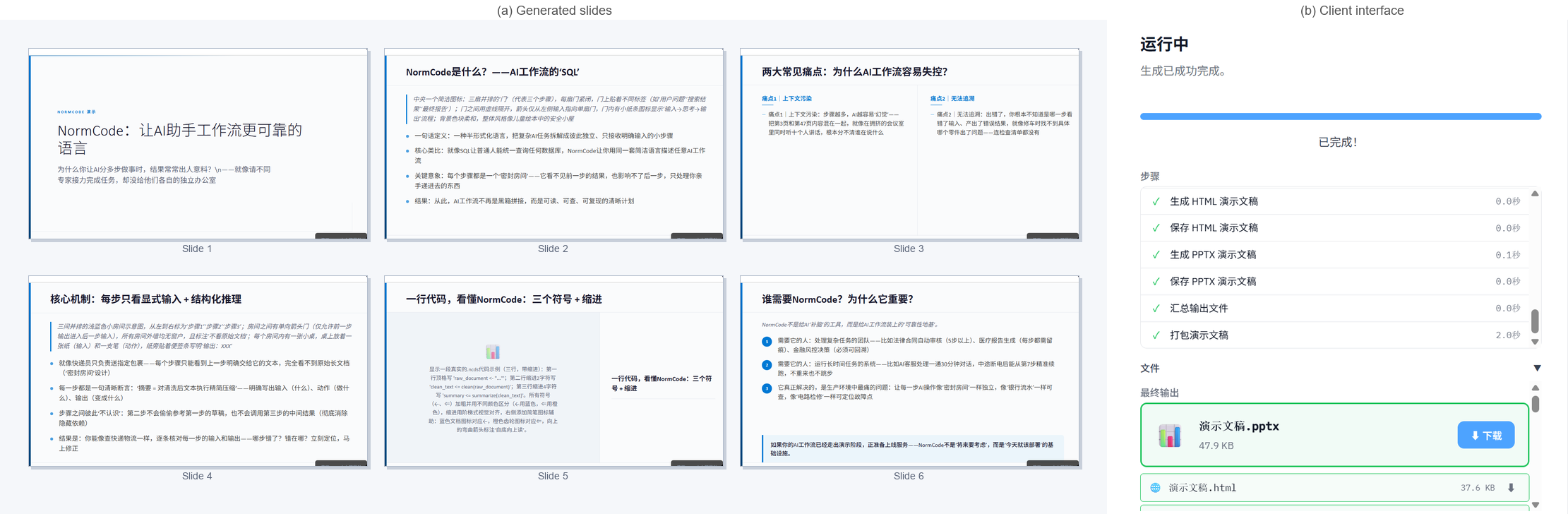}
  \caption{PPT Generation plan output. (a)~Six slides from a Chinese-language
    run; each slide is a Level~1 checkpoint. (b)~Client interface showing
    step-by-step progress and downloadable \texttt{.pptx}/\texttt{.html} output.}
  \label{fig:ppt}
\end{figure}

The entire plan is authored in Chinese --- every concept name
(\begin{CJK}{UTF8}{gbsn}\textit{分析}\end{CJK},
\begin{CJK}{UTF8}{gbsn}\textit{大纲}\end{CJK},
\begin{CJK}{UTF8}{gbsn}\textit{幻灯片内容}\end{CJK}, etc.) is a Chinese token ---
demonstrating full language agnosticism (Fig.~\ref{fig:ppt}).
The plan is \textbf{deployed as a server-side workflow}: clients (web wizard, CLI,
or desktop GUI) submit a topic and reference files and observe each step completing
in real time; on completion the client receives downloadable \texttt{.pptx} and
\texttt{.html} outputs. Per-slide timing shows content generation at ${\sim}4$\,s
and HTML rendering at ${\sim}12$--$17$\,s, while load, save, and collect nodes run
in under 0.2\,s --- confirming that 60--70\% of nodes are syntactic across all plans.

\subsection{Case Study 2: Code Assistant --- A ReAct Plan as a Level~2 Case}

\texttt{code\_assistant.ncds} (Fig.~\ref{fig:plans}, right) is a Level~2 case encoding
a ReAct-loop software-engineering workflow \cite{yao2023}. Each cycle, Explore
\textit{observes} (parallel file reads, zero LLM cost), Plan \textit{reasons} (one LLM
call over scoped exploration results), Implement \textit{acts} (surgical edits with an
inner lint-fix loop), and Verify checks results. This case most clearly illustrates the
scope rule: Implement receives the change plan, not the raw file contents from Explore.
That separation makes the evidence for C1 especially concrete, because the operator can
inspect the exact inputs passed into each stage rather than a merged conversational
history. It also supports C3: revising a planning or implementation checkpoint re-runs
only the downstream stages that depend on it. Every run generates a Level~1 case base,
so each iteration of the ReAct loop becomes inspectable and revisable in Canvas
(Sect.~\ref{sec:canvas}).

Table~\ref{tab:cases} summarizes all four deployed plans and their operational status
at the time of writing.

\begin{table}[!htbp]
  \centering
  \scriptsize
  \setlength{\tabcolsep}{3pt}
  \caption{Four deployed NormCode plans: workflow class, scale, and operational status.}
  \label{tab:cases}
  \begin{tabular}{p{1.7cm}p{2.4cm}p{2.6cm}p{4.4cm}}
    \toprule
    Plan & Workflow and scale & Control-flow pattern & Operational status \\
    \midrule
    PPT Generation & Slide-deck generation; ${\sim}30$ inferences; 6 layouts, 14 component types & Nested dual-output loop (HTML + PPTX per slide) & Runs on \texttt{qwen-turbo}/\texttt{qwen-plus} commercial APIs with negligible cost; produces a full presentation at ${\sim}40$\,s per slide. \\
    Code Assistant & Software-engineering workflow; ${\sim}40$ inferences & ReAct outer loop with parallel Explore & Executes up to 10-step reasoning cycles to carry out complex code editing tasks with scoped explore, plan, implement, and verify stages. \\
    NC~Compilations & Self-hosted compilation; ${\sim}25$ inferences & Sequential pipeline (Derivation $\rightarrow$ Activation) & With \texttt{qwen-plus}/\texttt{GLM-4.7}, produces within ${\sim}15$ steps executable plans for report writing, paper drafting, and contract review that run directly as new NormCode plans. \\
    Canvas Assistant & Built-in chat controller; ${\sim}20$ inferences & Interactive command-dispatch loop & Handles a wide range of canvas tasks; when connected to an AI code editor (Cursor, Claude Code), enables automated plan debugging by overtaking the canvas interface on the user's behalf. \\
    \bottomrule
  \end{tabular}
\end{table}

\subsection{Ecosystem Status}

In CBR, individual case quality improves as the case base grows. The same dynamic
applies at both levels of NormCode: every run deposits new Level~1 cases, and each
Level~2 plan fills a complementary role that reinforces the others.
NC~Compilations acts as the ecosystem's \textit{plan producer}, converting
natural-language task descriptions into executable plans within ${\sim}15$ steps.
The Code Assistant serves as an \textit{agentic explorer}, carrying out multi-step
software-engineering tasks with scoped explore, plan, implement, and verify stages;
each cycle enriches the case base with reusable intermediate states. The Canvas
Assistant functions as the \textit{plan debugger}: connected to an external AI code
editor such as Cursor \cite{cursor2024} or Claude Code \cite{claudecode2026}, it
overtakes the canvas to inspect checkpoints, diagnose failures, and apply fixes ---
automating the Level~1 CBR cycle for any plan.

These roles form a closed loop: NC~Compilations produces plans; those plans
accumulate Level~1 cases as they run; the Canvas Assistant debugs failing runs;
the Code Assistant extends the platform. Each plan's quality grows with the
ecosystem's accumulated experience, realizing cumulative learning at system scale.

\section{Discussion}
\label{sec:discussion}

\begin{figure}[!htbp]
  \centering
  \includegraphics[width=0.92\textwidth]{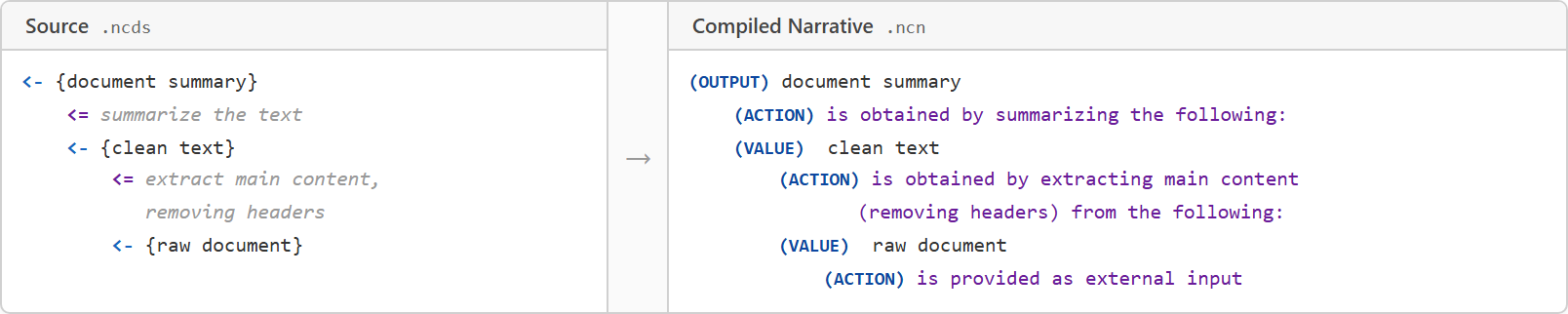}
  \caption{The \texttt{.ncn} narrative compiled from a \texttt{.ncds} source.
    Left: three-marker source. Right: deterministic plain-prose rendering used
    for pre-execution review~(C2). A domain expert reads the narrative to verify
    that the plan's logic matches intent before any LLM call.}
  \label{fig:ncn}
\end{figure}

\textbf{Language as the substrate for sustainable improvement.}
C1, C2, and C3 are structural consequences of the workflow language, not design
choices about the Canvas App --- and each directly enables a dimension of cumulative
improvement. C1~(failure localization) follows from Local Addressability: every step
has an exact flow index and exactly declared inputs, so every run deposits concrete
Level~1 cases that are immediately addressable for retrieval, reuse, and revision
without operator intervention. C2~(pre-execution review) follows from Interpretability
+ Enforceability: the \texttt{.ncn} is human-readable and \textit{exactly} what
executes (Fig.~\ref{fig:ncn}), enabling Level~2 case revision to be reviewed before
it generates any new Level~1 cases. C3~(selective re-execution) follows from Local Addressability +
Enforceability: the scope rule determines which downstream nodes depend on any revised
step, and the orchestrator enforces that boundary --- so a Level~2 case revision
propagates precisely, and a Level~1 fork re-executes only what is necessary.
Together, C1--C3 realize the CBR cycle at both levels: every run accumulates
experience (Level~1), every domain revision extends the abstract case library
(Level~2), and the compilation pipeline --- itself a Level~2 case --- improves through
the same cycle recursively \cite{bergmann2002,muller2015b}. Existing LLM orchestration
tools may expose similar checkpointing or graph-authoring affordances, but to our
knowledge they do not guarantee C1--C3 by construction without an equivalent scoped
intermediate representation satisfying the properties of Sect.~\ref{sec:normcode}.

Accordingly, our claims concern inspectability, revisability, and bounded
re-execution in multi-step LLM workflows, not universal superiority of final-task
quality.

\subsection{Limitations}

\textbf{Operational limitations.} Case retrieval currently requires operator judgment to
select the appropriate fork target; learned similarity measures would lower this barrier.
The case base grows without bound --- competence-based pruning is a natural next step.
Semantic checkpoints (LLM outputs) are bounded-stochastic: same input produces the same
output distribution, not a fixed value. Checkpoints whose tensors contain perceptual
signs are structurally self-contained, but reuse requires the referenced external
resources to be accessible; the case encodes \textit{what was referenced}, not its
content.

\textbf{Compilation scope.} NC~Compilations reliably produces executable plans for
tasks with clear sequential structure (document analysis, report generation, contract
review). Plans with complex domain logic or rich environment interaction (e.g.,
PPT~Generation's dual-format output loop) require human or code-assistant authoring;
the boundary is task complexity, not a language limitation.

\textbf{Evaluation scope.} Comprehensive quality evaluation would require a
combinatorial test matrix ($|L_2| \times |L_1|$ cells) where outcome quality also
depends on LLM model, input domain, and subjective judgment. The case studies
(Sect.~\ref{sec:casestudies}) argue for Level~2 generalizability --- that structural
properties hold across the four production plans and that C1--C3 (failure localization,
pre-execution review, selective re-execution) are realized --- rather than for absolute
output quality, which varies with domain, model choice, and subjective criteria.

\section{Related Work}
\label{sec:related}

\textbf{CBR and LLMs.} Aamodt \& Plaza \cite{aamodt1994} give the canonical CBR cycle; Bergmann \cite{bergmann2002}
and Minor et al.\ \cite{minor2014} establish POCBR as the field where cases capture
workflow knowledge. Within POCBR, Bergmann \& Gil \cite{bergmann2014} address retrieval
via structural similarity; Muller \& Bergmann \cite{muller2015a} introduce partial-workflow
query languages. Muller \& Bergmann \cite{muller2015b} treat abstract workflow patterns as case objects
--- the closest antecedent to NormCode's Level~2. Rubin et al.\ \cite{rubin2022} retrieve few-shot prompts; Wilkerson \& Leake
\cite{wilkerson2024} implement CBR components with LLMs; Lenz et al.\ \cite{lenz2025}
evaluate LLMs as similarity assessors. These works use CBR to improve individual LLM
calls. Guo et al.\ \cite{guo2024} (DS-Agent) is the most directly related: CBR
iteratively revises multi-step data-science plans. DS-Agent's cases are narrative plans
without enforcement guarantees; NormCode's checkpoints are structurally isolated by
compiler and runtime --- the precondition for reliable retrieval and revision
(Sect.~\ref{sec:background}).

\textbf{LLM Workflow Tools.} LangGraph, PromptFlow, and LangFlow provide checkpointing and graph-based authoring;
AutoGen \cite{wu2023} supports multi-agent orchestration. All pass state through
accumulated conversation histories or prompt templates, providing no structural
isolation guarantee. As a consequence, checkpoints embed implicit shared state, failures
are non-localizable, and case reuse may not reproduce the original data-dependency
behavior --- the three conditions in Sect.~\ref{sec:background}.

\textbf{LLM Agents and ReAct.} Yao et al.\ \cite{yao2023} propose ReAct (observe--reason--act); MetaGPT \cite{hong2024} encodes
software-engineering SOPs as prompt sequences; SWE-agent \cite{yang2024} introduces an
agent-computer interface. Deployed agentic coding tools --- Cursor \cite{cursor2024},
Claude Code \cite{claudecode2026}, and OpenClaw \cite{openclaw2026} --- implement
persistent observe-reason-act loops at production scale. NormCode expresses these
patterns as compiled plans: each loop iteration is a scoped node with a flow index and
automatic checkpoint. In standard ReAct the reasoning-action chain is an implicit
continuation with no structural transparency; AgentStepper \cite{agentstepper2026}
retrofits interactive debugging for this reason. NormCode aims to make these properties
structural from the start: every agent step is a Level~1 case, inspectable under
failure localization (C1) and revisable under selective re-execution (C3).

\newpage
\begin{credits}
\subsubsection{\ackname} This work was conducted at Psylens AI (Guangzhou Psylens Technology Studio), whose mission is to build auditable AI for transparent and controllable development --- making human values clearly expressible, recordable, and executable as sustainable technical assets. More information at \url{https://psylensai.com}. We thank the NormCode Canvas user community for workflow feedback that informed the case studies.

\subsubsection{Declaration on Generative AI}
During the preparation of this work, the author(s) used AI-assisted writing tools for
drafting and editing. After using these tools, the author(s) reviewed and edited the
content as needed and take(s) full responsibility for the publication's content.
\end{credits}

\bibliographystyle{splncs04}
\bibliography{bibliography}

@article{aamodt1994,
  author    = {Aamodt, Agnar and Plaza, Enric},
  title     = {Case-Based Reasoning: Foundational Issues, Methodological
               Variations, and System Approaches},
  journal   = {AI Communications},
  volume    = {7},
  number    = {1},
  pages     = {39--59},
  year      = {1994},
  publisher = {IOS Press},
  doi       = {10.3233/AIC-1994-7104}
}

@book{kolodner1993,
  author    = {Kolodner, Janet},
  title     = {Case-Based Reasoning},
  publisher = {Morgan Kaufmann},
  address   = {San Mateo, CA},
  year      = {1993},
  doi       = {10.1016/C2013-0-07426-7}
}

@book{bergmann2002,
  author    = {Bergmann, Ralph},
  title     = {Experience Management: Foundations, Development Methodology,
               and Internet-Based Applications},
  series    = {Lecture Notes in Computer Science},
  volume    = {2432},
  publisher = {Springer},
  address   = {Berlin, Heidelberg},
  year      = {2002},
  doi       = {10.1007/3-540-46237-4}
}

@article{bergmann2014,
  author    = {Bergmann, Ralph and Gil, Yolanda},
  title     = {Similarity Assessment and Efficient Retrieval of Semantic
               Workflows},
  journal   = {Information Systems},
  volume    = {40},
  pages     = {115--127},
  year      = {2014},
  publisher = {Elsevier},
  doi       = {10.1016/j.is.2012.11.004}
}

@article{minor2014,
  author    = {Minor, Mirjam and Montani, Stefania and Recio-Garcia, Juan A.},
  title     = {Process-Oriented Case-Based Reasoning},
  journal   = {Information Systems},
  volume    = {40},
  pages     = {103--105},
  year      = {2014},
  publisher = {Elsevier},
  doi       = {10.1016/j.is.2013.11.002}
}

@inproceedings{muller2015a,
  author    = {Muller, Gilles and Bergmann, Ralph},
  title     = {A Workflow-Based Query Language for Semantic Process Retrieval},
  booktitle = {Proceedings of the LWA 2015 Workshops},
  series    = {CEUR Workshop Proceedings},
  volume    = {1458},
  pages     = {247--255},
  year      = {2015},
  publisher = {CEUR-WS.org},
  url       = {https://ceur-ws.org/Vol-1458/}
}

@inproceedings{muller2015b,
  author    = {Muller, Gilles and Bergmann, Ralph},
  title     = {Generalized Workflow Cases: Learning Workflow-Based Case
               Adaptations},
  booktitle = {Proceedings of the 28th International FLAIRS Conference
               (FLAIRS-28)},
  year      = {2015},
  publisher = {AAAI Press}
}

@inproceedings{rubin2022,
  author    = {Rubin, Ohad and Herzig, Jonathan and Berant, Jonathan},
  title     = {Learning To Retrieve Prompts for In-Context Learning},
  booktitle = {Proceedings of the 2022 Conference of the North American
               Chapter of the Association for Computational Linguistics:
               Human Language Technologies (NAACL-HLT 2022)},
  pages     = {2655--2671},
  publisher = {Association for Computational Linguistics},
  address   = {Seattle, WA},
  year      = {2022},
  doi       = {10.18653/v1/2022.naacl-main.191}
}

@inproceedings{wilkerson2024,
  author    = {Wilkerson, Jacob and Leake, David B.},
  title     = {Large Language Models as Components in Case-Based Reasoning Systems},
  booktitle = {Case-Based Reasoning Research and Development (ICCBR 2024)},
  series    = {Lecture Notes in Computer Science},
  publisher = {Springer},
  address   = {Cham},
  year      = {2024},
  doi       = {10.1007/978-3-031-63646-2}
}

@inproceedings{lenz2025,
  author    = {Lenz, Lukas and Hoffmann, Maximilian and Bergmann, Ralph},
  title     = {{LLsiM}: Using Large Language Models as Similarity Measure for
               Case-Based Reasoning},
  booktitle = {Case-Based Reasoning Research and Development (ICCBR 2025)},
  series    = {Lecture Notes in Computer Science},
  volume    = {15662},
  pages     = {126--141},
  publisher = {Springer},
  address   = {Cham},
  year      = {2025},
  doi       = {10.1007/978-3-031-63646-2_9}
}

@article{guo2024,
  author    = {Guo, Siyuan and Deng, Cheng and He, Ailing and Huang, Feilong
               and Tong, Hanghang and Huang, Jing},
  title     = {{DS-Agent}: Automated Data Science by Empowering Large Language
               Models with Case-Based Reasoning},
  journal   = {arXiv preprint arXiv:2402.17453},
  year      = {2024},
  url       = {https://arxiv.org/abs/2402.17453}
}

@inproceedings{yao2023,
  author    = {Yao, Shunyu and Zhao, Jeffrey and Yu, Dian and Du, Nan and
               Shafran, Izhak and Narasimhan, Karthik and Cao, Yuan},
  title     = {{ReAct}: Synergizing Reasoning and Acting in Language Models},
  booktitle = {Proceedings of the International Conference on Learning
               Representations (ICLR 2023)},
  year      = {2023},
  url       = {https://openreview.net/forum?id=WE_vluYUL-X}
}

@article{agentstepper2026,
  author    = {Hutter, Robert and Pradel, Michael},
  title     = {{AgentStepper}: Interactive Debugging of Software Development Agents},
  journal   = {arXiv preprint arXiv:2602.06593},
  year      = {2026},
  url       = {https://arxiv.org/abs/2602.06593}
}

@inproceedings{hong2024,
  author    = {Hong, Sirui and Zhuge, Mingchen and Chen, Jonathan and Zheng, Xiawu
               and Cheng, Yuheng and Zhang, Ceyao and Wang, Jinlin and Wang, Zili
               and Yau, Steven Ka Shing and Lin, Zijuan and Zhou, Liyang and
               Ran, Chenyu and Xiao, Lingfeng and Wu, Chenglin and
               Schmidhuber, J{\"u}rgen},
  title     = {{MetaGPT}: Meta Programming for a Multi-Agent Collaborative Framework},
  booktitle = {Proceedings of the International Conference on Learning
               Representations (ICLR 2024)},
  year      = {2024},
  url       = {https://openreview.net/forum?id=VtmBAGCN7o}
}

@article{yang2024,
  author    = {Yang, John and Jimenez, Carlos E. and Wettig, Alexander and
               Lieret, Kilian and Yao, Shunyu and Narasimhan, Karthik and Press, Ofir},
  title     = {{SWE-agent}: Agent-Computer Interfaces Enable Automated Software
               Engineering},
  journal   = {arXiv preprint arXiv:2405.15793},
  year      = {2024},
  url       = {https://arxiv.org/abs/2405.15793}
}

@misc{cursor2024,
  author       = {{Anysphere}},
  title        = {{Cursor}: The {AI} Code Editor},
  howpublished = {Software},
  year         = {2024},
  url          = {https://www.cursor.com}
}

@misc{claudecode2026,
  author       = {{Anthropic}},
  title        = {{Claude Code}: Agentic Coding in the Terminal},
  howpublished = {Software},
  year         = {2025},
  url          = {https://docs.anthropic.com/en/docs/claude-code}
}

@misc{openclaw2026,
  author       = {{OpenClaw}},
  title        = {{OpenClaw}: Open-Source Agentic Coding Assistant},
  howpublished = {Software},
  year         = {2026},
  url          = {https://github.com/openclaw-ai/openclaw}
}

@misc{chase2022,
  author    = {Chase, Harrison},
  title     = {{LangChain}},
  howpublished = {GitHub repository},
  year      = {2022},
  url       = {https://github.com/langchain-ai/langchain}
}

@misc{langgraph2024,
  author    = {{LangChain AI}},
  title     = {{LangGraph}: Build Resilient Language Agents as Graphs},
  howpublished = {GitHub repository},
  year      = {2024},
  url       = {https://github.com/langchain-ai/langgraph}
}

@misc{microsoft2023,
  author    = {{Microsoft}},
  title     = {{PromptFlow}: Build High-Quality LLM Apps},
  howpublished = {GitHub repository},
  year      = {2023},
  url       = {https://github.com/microsoft/promptflow}
}

@article{wu2023,
  author    = {Wu, Qingyun and Bansal, Gagan and Zhang, Jieyu and Wu, Yiran
               and Li, Beibin and Zhu, Erkang and Jiang, Li and Zhang, Xiaoyun
               and Zhang, Shaokun and Liu, Jiale and Awadallah, Ahmed H.
               and White, Ryen W. and Burger, Doug and Wang, Chi},
  title     = {{AutoGen}: Enabling Next-Gen LLM Applications via Multi-Agent
               Conversation},
  journal   = {arXiv preprint arXiv:2308.08155},
  year      = {2023},
  url       = {https://arxiv.org/abs/2308.08155}
}

@inproceedings{ouyang2022,
  author    = {Ouyang, Long and Wu, Jeff and Jiang, Xu and Almeida, Diogo
               and Wainwright, Carroll L. and Mishkin, Pamela and Zhang, Chong
               and Agarwal, Sandhini and Slama, Katarina and Ray, Alex and
               Schulman, John and Hilton, Jacob and Kelton, Fraser and Miller,
               Luke and Simens, Maddie and Askell, Amanda and Welinder, Peter
               and Christiano, Paul and Leike, Jan and Lowe, Ryan},
  title     = {Training Language Models to Follow Instructions with Human
               Feedback},
  booktitle = {Advances in Neural Information Processing Systems (NeurIPS 2022)},
  volume    = {35},
  pages     = {27730--27744},
  year      = {2022},
  publisher = {Curran Associates, Inc.},
  url       = {https://proceedings.neurips.cc/paper_files/paper/2022/hash/b1efde53be364a73914f58805a001731-Abstract-Conference.html}
}

@article{bai2022,
  author    = {Bai, Yuntao and Jones, Andy and Ndousse, Kamal and Askell, Amanda
               and Chen, Anna and DasSarma, Nova and Drain, Dawn and Fort,
               Stanislav and Ganguli, Deep and Henighan, Tom and Joseph, Nicholas
               and Kadavath, Saurav and Kernion, Jackson and Conerly, Tom and
               El-Showk, Sheer and Elhage, Nelson and Hatfield-Dodds, Zac and
               Hernandez, Danny and Hume, Tristan and Johnston, Scott and Kravec,
               Shauna and Lovitt, Liane and Nanda, Neel and Olsson, Catherine
               and Amodei, Dario and Brown, Tom and Clark, Jack and McCandlish,
               Sam and Olah, Chris and Mann, Ben and Kaplan, Jared},
  title     = {Constitutional {AI}: Harmlessness from {AI} Feedback},
  journal   = {arXiv preprint arXiv:2212.08073},
  year      = {2022},
  url       = {https://arxiv.org/abs/2212.08073}
}

@misc{guan2026normcodesemiformallanguageauditable,
  title     = {{NormCode}: A Semi-Formal Language for Auditable {AI} Planning},
  author    = {Xin Guan and Yunshan Li and Zekun Wu and Ruibo Zhang},
  year      = {2026},
  eprint    = {2512.10563},
  archivePrefix = {arXiv},
  primaryClass  = {cs.AI},
  url       = {https://arxiv.org/abs/2512.10563},
}

\end{document}